%% file: pushkarev.tex
\documentclass{evn2004}
\usepackage{txfonts}
\usepackage{graphicx}
\begin{document}
   \title{LFVN observations of active galactic nuclei}
   

   \author{A.~Pushkarev\inst{1}, 
   I.~Molotov\inst{1}, 
   M.~Nechaeva\inst{2}, Yu.~Gorshenkov\inst{3}, 
   G.~Tuccari\inst{4}, C. Stanghellini\inst{4}, X.~Hong\inst{5}, X. Liu\inst{6}, 
   J.~Quick\inst{7}, \and S.~Dougherty\inst{8}
          }

\institute{Central Atronomical Observatory, Pulkovo, Pulkovskoe sh. 65/1, 196140 St.-Petersburg, Russia
\and
Research Institute, B. Pecherskaya 25, 603950 N. Novgorod, Russia
\and
Special Research Bureau, MPEI, Krasnokazarmennaya 14, 111250 Moscow, Russia
\and
Istituto di Radioastronomia, Contrada Renna Bassa, 96017 Noto, Italy
\and
Shanghai Astronomical Observatory, NAO CAS, Nandan road 80, 200030 Shanghai, China
\and
Urumqi Astronomical Observatory, NAO CAS, S. Beijing Road 40, 830011 Urumqi, China
\and
Hartebeesthoek Radio Astronomy Observatory, P.O. Box 443, 1740 Krugersdorp, South Africa
\and
Dominion Radio Astrophysical Observatory, P.O. Box 248, V2A 6K3 Penticton B.C., Canada
             }

   \abstract{In November 1999 we carried out VLBI observations of several quasars
and BL Lacertae objects at 1.66~GHz.  Six antennas participated in the
experiment (Bear Lakes, Svetloe, Pushchino, Noto, HartRAO, and Seshan). The
results for six sources
(0420+022, 0420$-$014, 1308+326, 1345+125, 1803+784, and DA~193) are presented
and discussed.
  }
   \authorrunning{A. Pushkarev, I. Molotov, M. Nechaeva et al.}
   \titlerunning{LFVN observations of AGN}
   \maketitle
%

\section{Introduction}

LFVN project was started in 1996 (Molotov et al. \cite{molotov02})
having the purpose to arrange the
international VLBI cooperation with participation of former Soviet
Union radio telescopes. 

At first stages, LFVN was developing dynamically and successfully. 
13 antennas in different countries were equipped with new receiving-recording 
radio astronomy apparatuses: Bear Lakes RT-64, Pushchino RT-22, Zimenki RT-15,
St. Pustyn RT-14 (Russia), Evpatoria RT-70 and Simeiz RT-22 (Ukraine), 
Ventspils RT-32 (Latvia), Noto RT-32 (Italy), Torun RT-14 (Poland), Ooty 
500x30 parabolic cylinder of ORT and Pune RT-45 of GMRT (India), Urumqi 
RT-25 and Shanghai RT-25 (China). 

The 18 VLBI experiments were carried out using various combinations of 
radio telescopes in Canada, China, England, India, Italy, Japan, Latvia, 
Poland, Russia, South Africa, Ukraine and USA. LFVN had two main directions 
of works: (i) a subsystem with use of obsolete, but very cheap in operations, 
Mk-2 recording terminal for investigations of solar wind, solar spikes and 
radar research of Earth group planets, close asteroids and space debris 
objects; (ii) a subsystem with use of more modern S2 Canadian recording 
terminal (Cannon et al. \cite{cannon97}) for the investigations of 
active galactic nuclei (AGN), solar wind, 
OH-masers, active stars. At first, LFVN experiments were processed by 
Block II JPL/Caltech Mk-2 correlator in USA, then by the Penticton DRAO S2 
correlator in Canada (Carlson et al. \cite{carlson99}). 
Step by step, LFVN acquired all necessary signs of 
VLBI network - technical, scheduling and post-processing groups, ``own'' Mk-2 
correlator NIRFI-3 in N. Novgorod, Russia. LFVN results on AGN, solar wind, 
spikes, radar researches, and technical developments were published in more 
than 100 papers in open scientific literature.  The fully steerable NIS radio 
telescopes participated in LFVN experiments and few NIS antennas had first 
VLBI fringes exactly under this work. Two largest NIS radio 
telescopes: Bear Lakes RT-64 and Evpatoria RT-70 are operated under LFVN activity 
mainly. St. Pustyn RT-14 is operated only under LFVN activity. LFVN is 
a single European instrument that regularly carries out the VLBI radar 
observations of the Solar system bodies. 

The  three  VLBI sessions with participation of six telescopes each
were mainly devoted to the research of AGN. The sample of GPS sources 
and BL Lacs were observed. 


\section{Observations}
The observations (INTAS~99.4 experiment) were made in November 1999 (1999.91) at 1.66~GHz
using a global VLBI array of six antennas listed in Table~\ref{antennas}. 
The north-south resolution provided by these observations is substantially improved due
to the use of the southern antenna in South Africa. The data were
recorded using the S2 system, and all data were subsequently correlated
using the Penticton S2 correlator. The data were calibrated and imaged
in the NRAO AIPS package using the standard technique.
In addition to the results reported below on the observations of
active galactic nuclei, the experiment had some more targets to explore:
solar wind study, an OH-maser monitoring, and observations of a radiostar
Lambda And. 


\begin{table}
      \caption[]{Antennas and their characteritics at 1.66~GHz.}
         \label{antennas}
     $$
         \begin{array}{p{0.4\linewidth}cr}
            \hline
            \noalign{\smallskip}
            Radio  telescope &  \mathrm{Diameter} &  \mathrm{SEFD}^{\mathrm{a}} \\
	                     &  \mathrm{(m)}      &  \mathrm{(Jy)} \\
            \noalign{\smallskip}
            \hline
            \noalign{\smallskip}
Svetloe (Russia)       &  32      &   394 \\
Bear Lakes (Russia)    &  64      &   156 \\
Pushchino (Russia)     &  22      &  1585 \\
HartRAO (South Africa) &  26      &   515 \\
Noto (Italy)           &  32      &  1070 \\
SESHAN25 (China)       &  25      &  1250 \\            
            \noalign{\smallskip}
            \hline
         \end{array}
     $$ 
\begin{list}{}{}
\item[$^{\mathrm{a}}$] System equivalent flux density
\end{list}
   \end{table}
  
\section{Results}

Results for the six sources are presented below.
Models for the source structures were derived modelling
the complex $I$ visibilities that come from the hybrid mapping process with 
circular Gaussian components, as described by Roberts et al.\ (\cite{rgw87}). 
The model fits are shown in Table~\ref{models}. We denote distance 
from the core by $r$ and the VLBI jet direction by $\varphi$. 
The errors of the separations of jet components from the core are
formal $1\sigma$ errors, corresponding to an increase in the best-fitting $\chi^2$ by unity.
The smallest of these formal errors underestimate the actual errrors;
realisticaly, the smallest $1\sigma$ errors are probably no less than 0.3~mas at at 1.6~GHz.

\begin{figure}
   \centering
  \includegraphics[width=8cm]{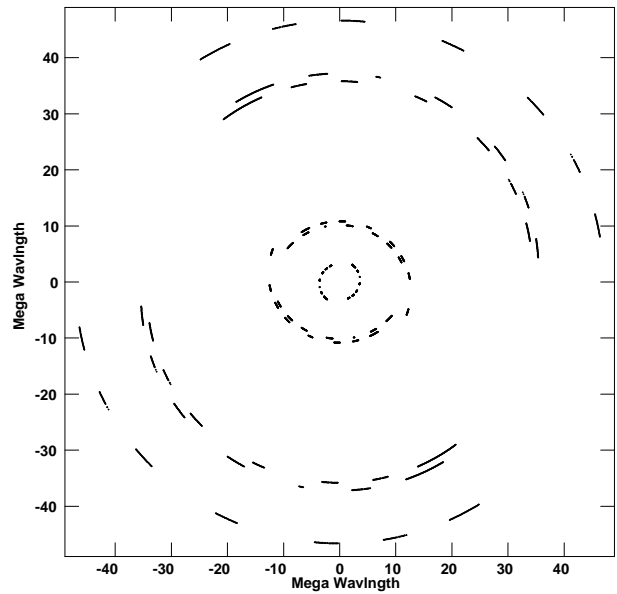}
     \caption{Coverage of the {\it u-v} plane at 1.66~GHz for LFVN observations of 1803+784 at epoch 1999.91.
        \label{1803uv}
        }
  \end{figure}

\begin{figure}
   \centering
  \includegraphics[width=8cm]{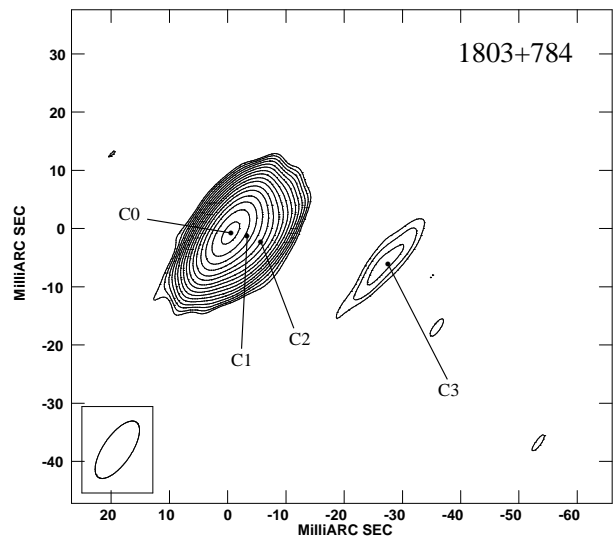}
     \caption{LFVN image of 1803+784 at 1.66~GHz with bottom contour at 0.5\% of the peak of 1.48~Jy/beam.
     Positive contour levels increase by a factor of $\sqrt2$.
        \label{1803map}
        }
  \end{figure}

Fig.~\ref{1803uv} shows {\it u-v} plane coverage of BL Lac object 1803+784. VLBI image of this
source is presented on Fig.~\ref{1803map}. 

Using our model and a model taken by Hong et al. (\cite{hong99}) for variable and strongly 
polarised quasar 0420$-$022 (z=0.915) at epoch 1995.83 we were able to estimate an apparent 
speed of the jet component located at 1.78~mas from the VLBI core. 
Assuming $H_0=70h$~km~s$^{-1}$ Mpc$^{-1}$ and $q_0=0.5$ we obtained $\beta_{app}=1.3h^{-1}$.

\begin{table}
\caption{Source models.}
\label{models}
\begin{tabular}{rrrc}
\noalign{\smallskip}
\hline
\noalign{\smallskip}
 $I\pm\sigma_I$,& $r\pm\sigma_r$, & $\varphi\pm\sigma_\varphi$, & $\theta\pm\sigma_\theta$,\\
  mJy          & mas &   deg                     & mas\\
\noalign{\smallskip}
\hline
\noalign{\smallskip}
\multicolumn{4}{c}{0420$-$022}\\
\noalign{\smallskip}
\hline
\noalign{\smallskip}
 $1309 \pm 42$ &  \dots           & \dots          & $<0.5         $ \\
  $110 \pm 23$ & $ 1.78 \pm 0.12$ & $ 159 \pm 5.5$ & $<0.5         $ \\
   $30 \pm 4$  & $ 7.06 \pm 0.22$ & $-176 \pm 0.7$ & $<0.5         $ \\
  $ 20 \pm 5$  & $12.59 \pm 0.38$ & $-176 \pm 0.8$ & $<0.5         $ \\
\noalign{\smallskip}	   
\hline
\noalign{\smallskip}
\multicolumn{4}{c}{0420$+$022}\\
\noalign{\smallskip}
\hline
\noalign{\smallskip}
 $772 \pm 28$ &  \dots           &  \dots         & $<0.5         $ \\
  $52 \pm 9$  & $ 1.78 \pm 0.14$ &  $-41 \pm 4.6$ & $<0.5         $ \\
  $61 \pm 8$  & $ 2.31 \pm 0.21$ &  $-86 \pm 4.8$ & $0.63 \pm 0.25$ \\
  $73 \pm 8$  & $ 4.53 \pm 0.25$ &  $-93 \pm 3.3$ & $<0.64        $ \\
\noalign{\smallskip}
\hline
\noalign{\smallskip}
\multicolumn{4}{c}{DA\,193}\\
\noalign{\smallskip}
\hline
\noalign{\smallskip}
 $1872 \pm 8$  &  \dots           &  \dots         & $0.55 \pm 0.05$ \\
\noalign{\smallskip}
\hline
\noalign{\smallskip}
\multicolumn{4}{c}{1308$+$326}\\
\noalign{\smallskip}
\hline
\noalign{\smallskip}
 $1389 \pm 44$ &  \dots           &  \dots         & $1.11 \pm 0.10$ \\
  $925 \pm 68$ & $ 2.33 \pm 0.13$ & $  35 \pm 3.1$ & $1.97 \pm 0.23$ \\
  $299 \pm 60$ & $ 4.99 \pm 0.15$ & $  54 \pm 1.8$ & $1.45 \pm 0.61$ \\
\noalign{\smallskip}	   
\hline
\noalign{\smallskip}
\multicolumn{4}{c}{1803$+$784}\\
\noalign{\smallskip}	   
\hline
\noalign{\smallskip}
 $1459 \pm 87$ &  \dots           &  \dots         & $1.10 \pm 0.26$ \\
   $80 \pm 16$ & $ 2.80 \pm 0.23$ & $-100 \pm 4.1$ & $0.89 \pm 0.31$ \\
  $322 \pm 31$ & $ 4.83 \pm 0.46$ & $-104 \pm 5.2$ & $1.99 \pm 0.88$ \\
   $36 \pm 9$  & $26.24 \pm 0.87$ & $-103 \pm 1.2$ & $1.18 \pm 0.35$ \\
\noalign{\smallskip}	   
\hline
\end{tabular}
\end{table}

%

%

\begin{acknowledgements}
This research was supported by INTAS 2001-0669. We thank the staff at the
participating observatories who made these observations possible.
\end{acknowledgements}

\end{document}